\documentclass[12pt]{article}
\usepackage{graphicx}
\usepackage{amssymb}
\usepackage{amscd}
\usepackage{amsmath}
\usepackage{appendix}
\usepackage{epstopdf}

\begin{document}
\begin{titlepage}

{\hbox to\hsize{\hfill March 2016 }}

\bigskip \vspace{3\baselineskip}

\begin{center}
{\bf \large
750 GeV Composite Axion as the LHC Diphoton Resonance }

\bigskip

\bigskip

{\bf Neil D. Barrie, Archil Kobakhidze, Matthew Talia and Lei Wu \\ }

\smallskip

{ \small \it
ARC Centre of Excellence for Particle Physics at the Terascale, \\
School of Physics, The University of Sydney, NSW 2006, Australia \\
E-mails: neil.barrie, archil.kobakhidze,matthew.talia, lei.wu1@sydney.edu.au
\\}

\bigskip

\bigskip

\bigskip

{\large \bf Abstract}

\end{center}
\noindent
We propose that the 750 GeV resonance, presumably observed in the early LHC Run 2 data, could be a heavy composite axion that results from condensation of a hypothetical quark in a high-colour representation of conventional QCD. The model, motivated by a recently proposed solution to the strong CP problem, is very economical and is essentially defined by the properties of the additional quark - its colour charge, hypercharge and mass. The axion mass and its coupling to two photons (via axial anomaly) can be computed in terms of these parameters. The axion is predominantly produced via photon fusion ($\gamma\gamma \to {\cal A}$) which is followed by $ Z $ vector boson fusion and associated production at the LHC. We find that the total diphoton cross section of the axion can be fitted with the observed excess. Combining the requirement on the cross-section, such that it reproduces the diphoton excess events, with the bounds on the total width ($\Gamma_{tot} \leqslant 45$ GeV), we obtain the effective coupling in the range $1.6\times 10^{-4}$ GeV$^{-1}\gtrsim C_{{\cal A}} \gtrsim 6.5\times 10^{-5}$ GeV$^{-1}$. Within this window of allowed couplings the model favours a narrow width resonance and $ y_{Q}^2 \sim \mathcal{O}(10)$. In addition, we observe that the associated production $q\bar{q} \to {\cal A}\gamma\to \gamma\gamma\gamma$ can potentially produce a sizeable number of three photon events at future LHC. However, the rare decay $Z\to\mathcal{A}^*\gamma \to \gamma\gamma\gamma$ is found to be too small to be probed at the LHC and $ e^{+}  e^{-} $ colliders.

\end{titlepage}


\section{Introduction}
There is preliminary evidence for a new diphoton resonance at $\sim 750$ GeV in the early LHC Run 2 data. ATLAS  reports an excess of 15 events (with detector efficiency 40\%) in four bins around 750 GeV in their 3.2 fb$^{-1}$ 13 TeV data \cite{ATLAS750}. This presumably implies a relatively large width for the putative resonance, fitted to $\Gamma=45$ GeV in one model template. The local (global) statistical significance of the observed excess is $3.9\sigma$ ($2.3\sigma$). CMS also sees an excess, which consists of $\sim 10$ events peaked at 760 GeV \cite{CMS750}. The CMS excess is consistent with the narrow width resonance at $2.6\sigma$ ($1.2\sigma$) local (global) significance.

The anomalous events are not accompanied by any significant missing energy, neither leptons or jets are seen. No resonances at invariant mass 750 GeV are seen in the Run 2 data in $ZZ$, $W^+W^-$, $t\bar{t}$ or $jj$ events and the diphoton resonance has not been observed by ATLAS in 8 TeV Run 1 data.
The $pp\to R\to \gamma\gamma$ cross section needed to explain the excess is estimated as \cite{Franceschini:2015kwy}:
\begin{eqnarray}
\label{1.1}
\sigma^{\rm ATLAS}(pp\to R \to \gamma\gamma)&\approx &(10\pm 3)~ {\rm fb} \\
\label{1.2}
\sigma^{\rm CMS}(pp\to R \to \gamma\gamma)&\approx & (6\pm 3)~ {\rm fb}
\end{eqnarray}
The 8 TeV Run 1 data is compatible with the 13 TeV Run 2 data at the level of $2\sigma$, providing the signal cross section grows by at least a factor of $\sim 4$-$5$. Together with the 8 TeV data, the diphoton excess contributing to the combined production rate is given by \cite{Buttazzo:2015txu}
\begin{equation}
\sigma(pp \to R \to \gamma\gamma) = (4.4\pm 1.1) ~\rm{fb}~.\label{excess}
\end{equation}

From the available data it is premature to draw any definite conclusion about the properties of the diphoton resonance, including its very existence. Nevertheless, assuming the observed excess is not a statistical fluke, the suggested large branching ratio of $R\to \gamma\gamma$ is strongly reminiscent of the composite neutral pion. The vast majority of composite models  discussed so far in relation to the LHC diphoton resonance (see, presumably, an incomplete list of Refs. \cite{Harigaya:2015ezk}-\cite{Franzosi:2016wtl}), assumed production of the resonance via gluon fusion. It has been pointed, however, that the excess can be accommodated in a more minimal scenario via photo-production (\cite{Fichet:2015vvy}-\cite{Harland-Lang:2016qjy}). Drell-Yan production $pp\to RR \to 4\gamma$ has also been discussed in \cite{Cline:2015msi,Huang:2015evq} within the effective field theory approach. Specific models relying on the photo-production mechanism can be found in \cite{Anchordoqui:2015jxc, Abel:2016pyc,  Ben-Dayan:2016gxw}.

In this paper we propose a very economical model where a pion-like heavy composite axion is responsible for the LHC diphoton excess. Theoretically the model is motivated by a recently proposed solution to the strong CP problem \cite{strongCP}, and involves high-colour (other than triplet) vector-like quark(s). The quark-antiquark condensate breaks the axial symmetry associated with the high-colour quark and results in a composite axion that couples predominantly to two photons. The axion mass and its coupling to two photons (via axial anomaly) can be computed essentially in terms of the colour charge, hypercharge and the mass of the high-colour quark. In what follows we describe the model in more details and study the diphoton resonance in this framework.

\section{High-colour quarks and the heavy composite axion}
The model we propose for the 750 GeV LHC resonance is extremely economical. In the minimal scenario we just add to the Standard Model a vector-like quark ${\mathcal{Q}}$ in some representation $\mathcal{R}$ of the $SU(3)$ colour gauge symmetry group. We also assume that it carries the hypercharge $y_{{\mathcal{Q}}}$ and is a weak isospin singlet. For the higher than colour triplet representation $\mathcal{R}$, there is a separate $U(1)_V\times U(1)_A$ global symmetry of strong interactions associated with ${\mathcal{Q}}$ quark number conservation and chiral phase rotation of ${\mathcal{Q}}$, respectively. The latter axial $U(1)_A$ symmetry is broken explicitly by a ${\mathcal{Q}}$ quark mass and the QCD and hypercharge anomalies. The divergence of the axial current density $J^{\mu}_A=\bar {\mathcal{Q}}\gamma^{\mu}\gamma^5{\mathcal{Q}}$ reads:
\begin{equation}
\partial_{\mu}J^{\mu}_A=2im_{\mathcal{Q}} \bar{\mathcal{Q}}\gamma^5\mathcal{Q}+
\frac{{\rm T}(\mathcal{R})\alpha_3}{4\pi}\epsilon^{\mu\nu\alpha\beta}G_{\mu\nu}^aG_{\alpha\beta}^a
+\frac{{\rm d}(\mathcal{R})y_{\mathcal{Q}}^2\alpha_1}{4\pi}\epsilon^{\mu\nu\alpha\beta}B_{\mu\nu}B_{\alpha\beta}~,
\label{2.1}
\end{equation}
where $G_{\mu\nu}^a$ and $B_{\mu\nu}$ are gluon and hypercharge field strength tensors, respectively; $\alpha_3$ is the colour SU(3) and $\alpha_1$ is the U(1)$_Y$ hypercharge fine-structure constants; ${\rm T}(\mathcal{R})$ is the Dynkin index for the SU(3) representation $\mathcal{R}$, which is expressed through the corresponding eigenvalue of the quadratic Casimir operator, ${\rm C}_2(\mathcal{R})$, and the dimension of the representation, ${\rm d}(\mathcal{R})$, as:
\begin{equation}
{\rm T}(\mathcal{R})=\frac{1}{8}{\rm C}_{2}(\mathcal{R}){\rm d}(\mathcal{R})~.
\label{2}
\end{equation}

Recall a separate axial symmetry in the sector of conventional light quarks, $q=(u,d)^{\rm T}$, is also anomalous:
\begin{equation}
\partial_{\mu}j^{\mu}_A=2im_{u} \bar u\gamma^5 u+2im_{d} \bar d\gamma^5 d
+\frac{\alpha_3}{4\pi}\epsilon^{\mu\nu\alpha\beta}G_{\mu\nu}^aG_{\alpha\beta}^a
+\frac{3(y_u^2+y_d^2)\alpha_1}{8\pi}\epsilon^{\mu\nu\alpha\beta}B_{\mu\nu}B_{\alpha\beta}~,
\label{2.2}
\end{equation}
where $j^{\mu}_A=\bar u\gamma^5 u+\bar d\gamma^5 d$ and $y_u=4/3$ and $y_d=-2/3$ are hypercharges of right-handed up and down quarks, respectively. The colour anomaly actually eliminates one of these axial symmetries and through the QCD instantons solves the $\eta$ mass problem. The symmetry which is colour anomaly-free results in the following N\"oether current density:
\begin{eqnarray}
\tilde J^{\mu}_A= j^{\mu}_A - {\rm T}(\mathcal{R})J^{\mu}_A~.
\label{2.3}
\end{eqnarray}
The corresponding charge is:
\begin{equation}
Q_5=\int d^3x~ \tilde J^{0}_A~.
\label{2.4}
\end{equation}

Just like in the case of light quarks, the conventional colour forces bound $\mathcal{Q}$ quarks into colourless hadrons. The composite meson  would then condense,
\begin{equation}
\langle 0 \vert \bar{\mathcal{Q}}\mathcal{Q} \vert 0\rangle = -cF_{\mathcal{A}}~,
\label{2.5}
\end{equation}
breaking the axial symmetry, generated by the charge, (\ref{2.4}), spontaneously [$c$ is a constant $\sim \mathcal{O}(1)$].
This breaking is accompanied by a composite pseudoscalar, $\mathcal{A} \sim \bar{\mathcal{Q}}\gamma^5 \mathcal{Q}$, the pseudo-Goldstone boson of the spontaneously broken approximate axial symmetry. With lack of imagination, we call it the heavy axion. We would like to associate this composite axion with the 750 GeV LHC resonance.

Due to the high-colour representation, the scale of axial symmetry breaking in $\mathcal{Q}$-sector is naturally hierarchically larger than the scale of the chiral symmetry breaking in the conventional quark sector, $F_{\mathcal{A}}\gg f_{\pi}\approx 130$ MeV. We also assume that $m_{\mathcal{Q}}\gg m_{u,d}$, therefore $\mathcal{Q}$-sector can be safely considered as being decoupled from the $q$-sector.  Following \cite{Marciano:1980zf}, we roughly estimate:
\begin{equation}
\mathcal{F}_{\mathcal{A}}\approx f_{\pi} \exp\left[\frac{2\pi}{7\alpha_3(\Lambda_{QCD})}\left(\frac{3}{4}{\rm C}_2(\mathcal{R})-1\right)\right]~,
\label{2.6}
\end{equation}
where $\Lambda_{\rm QCD}\sim 1$ GeV. One can see that $\mathcal{F}_{\mathcal{A}}$ is very sensitive to values of the strong coupling constant $\alpha_3$ at low energies. At the same time, determination of $\alpha_3$ at low energies contains significant uncertainties. For example, the estimate obtained from the charmonium fine stricture splitting is $\alpha_3(1~{\rm  GeV})\approx 0.38\pm 0.05$ \cite{Badalian:1999fq}, while extracting $\alpha_3$ from the hadronic decays of tau lepton gives $\alpha_3(1.7~{\rm GeV})\approx 0.331\pm 0.013$ \cite{Pich:2013lsa}. In our estimations below we allow $\alpha_3(\Lambda_{QCD})=0.3-0.5$. With this, to have a feel of numbers, we presented some estimates in Table 1.
\begin{table}

\begin{center}
\begin{tabular}{|l|l|l|l|}
\hline
Repr. & C$_2(\mathcal{R})$ & T$(\mathcal{R})$ & $\mathcal{F}_{\mathcal{A}}$, GeV \\
\hline
\hline
d($\mathcal{R}$)=6 & $10/3$ & $5/2$ & $2.0-12.0$ \\
\hline
d($\mathcal{R}$)=8 &  $3$ & $3$ & $1.3-6.0$ \\
\hline
d($\mathcal{R}$)=10 & $6$&  $15/2$ & $75-4590.0$ \\
\hline
d($\mathcal{R}$)=15 &  $28/3$ & $35/2$ &$(0.007-8.13)\cdot 10^6$ \\
\hline
\end{tabular}
\end{center}
\caption{\small Estimates of $\mathcal{F}_{\mathcal{A}}$ for various high-colour representations according to Eq. (\ref{2.6}). The strong coupling
is assumed in the range $\alpha_3 (\Lambda_{QCD})=0.3-0.5$. }
\end{table}


The mass of the axion can be computed using the standard current algebra technique. In the decoupling limit, Dashen's formula \cite{Dashen:1969eg} gives:
\begin{equation}
m_{\mathcal{A}}^2=-\frac{1}{\mathcal{F}_{\mathcal{A}}^2}\langle 0\vert\left[Q_5, \partial_{\mu}\tilde J^{\mu}_A \right] \vert 0\rangle \approx
-\frac{4}{\mathcal{F}_{\mathcal{A}}^2}\frac{m_{\mathcal{Q}}}{{\rm T}^2(\mathcal{R})}\langle 0 \vert \bar{\mathcal{Q}}\mathcal{Q} \vert 0\rangle \approx 4c m_{\mathcal{Q}}\mathcal{F}_{\mathcal{A}}~,
\label{2.7}
\end{equation}
Hence, the mass of the heavy axion, $m_{\mathcal{A}}\approx 750$ GeV, is defined (up to a constant $c$) by  the $\mathcal{Q}$ quark mass and the axion decay constant.

The heavy axion coupling to the hypercharge can be read off from the hypercharge anomaly of the current density Eq. (\ref{2.1}). Ignoring again the axion mixing with the light mesons we compute:
\begin{equation}
\mathcal{L}_{\mathcal{A}BB}= C_{\mathcal{A}} \epsilon^{\mu\nu\alpha\beta} \mathcal{A} B_{\mu\nu}B_{\alpha\beta}~,
\label{2.8}
\end{equation}
 with
 \begin{equation}
  C_{\mathcal{A}}= {\rm d}(\mathcal{R})\frac{y_{\mathcal{Q}}^2\alpha_1}{4\pi \mathcal{F}_{\mathcal{A}}}~.
\label{2.9}
 \end{equation}
 We observe here an enhancement by a factor ${\rm d}(\mathcal{R})$ due to the high-colour representation of $\mathcal{Q}$. Recalling that $B_{\mu\nu}=\cos\theta_{W} F_{\mu\nu}^{\gamma}- \sin \theta_W F^Z_{\mu\nu}$, we can extract from Eq. (\ref{2.9}) $\mathcal{A}\gamma\gamma$, $\mathcal{A}\gamma Z$ and $\mathcal{A}ZZ$ couplings, respectively:
  \begin{equation}
 C_{\gamma\gamma}=C_{\mathcal{A}}\cos^2\theta_{W}~,~~C_{\gamma Z}=-2\tan\theta_{W} C_{\gamma\gamma}~, C_{ZZ}=\tan^2\theta_{W} C_{\gamma\gamma}~,
\label{2.10}
 \end{equation}
where $\theta_W$ is the weak mixing angle, $\tan\theta_W\approx 0.55$.

The composite axion in principle decays to ordinary quarks and leptons through weak processes. They also may decay into light mesons as well as to exotic bound states of ordinary quarks and the high-colour quark. Hence, the total width of the axion can in principle be sizeable. We keep it as a free parameter in our numerical analysis, subject to the constraint $ \Gamma_{tot}\leq 45 $ GeV, where
\begin{equation}
\Gamma_{tot}=\Gamma_{\gamma\gamma}+\Gamma_{Z\gamma}+\Gamma_{ZZ}+\Delta\Gamma.
\end{equation}

We note that the collider phenomenology of a massive pseudoscalar field coupled to hypercharge topological density has been discussed long ago in \cite{Brustein:1999it,Elfgren:2000ch,Brustein:2000pd}

\section{Numerical Results and Discussions}

\begin{figure}[h]
	\centering
	\includegraphics[width=0.9\textwidth]{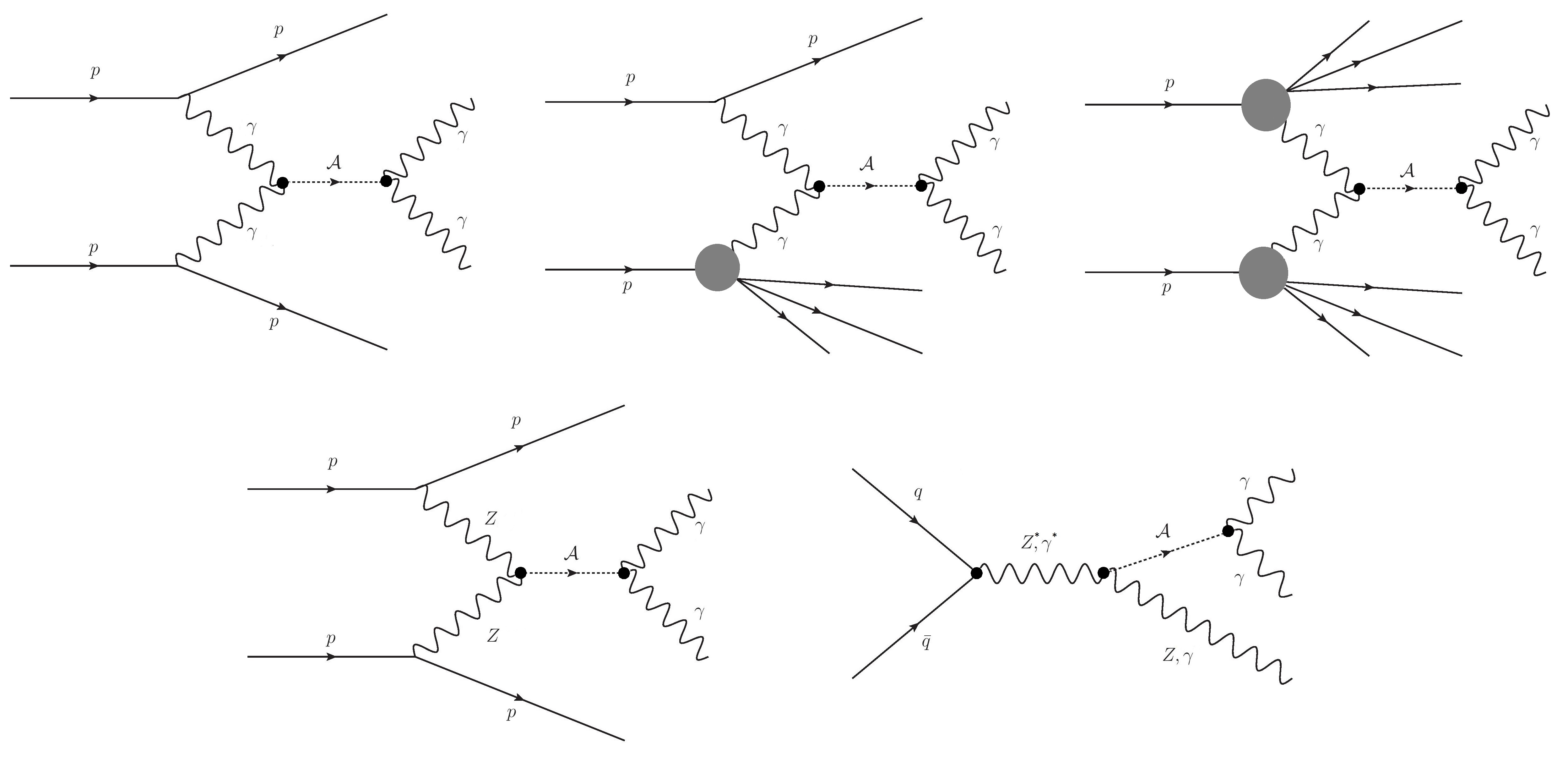}
	\caption{The Feynman diagrams for photon fusion (inelastic, semi-elastic and elastic scattering subprocesses from left to right in top panel), the weak vector boson fusion (left-lower panel) and associated production (right-lower panel) of the heavy axion ${\cal A}$.}
	\label{feyn}
\end{figure}
In our study, there are three subprocesses contributing to the heavy axion inclusive production rate, which are the photon fusion production ($ \gamma\gamma \to {\cal A}$), the $ Z $ vector boson fusion production ($qq \to {\cal A}qq$) and the associated production ($q\bar{q} \to {\cal A}\gamma/Z$). The corresponding Feynman diagrams are shown in Fig.\ref{feyn}. The contribution to the photon-fusion is dominated by the inelastic scattering, which is followed by the semi-elastic and elastic processes. The Lagrangian described by Eq. (\ref{2.8}) is implemented by using \texttt{FeynRules} \cite{Alloul:2013bka}. The cross sections of the above production processes are calculated with \texttt{Madgraph 5} \cite{Alwall:2014hca} at $\sqrt{s} = $ 13 TeV with \texttt{NN23LO1} PDFs \cite{Ball:2012cx}.

\begin{figure}[h]
	\centering
	\includegraphics[width=0.5\textwidth]{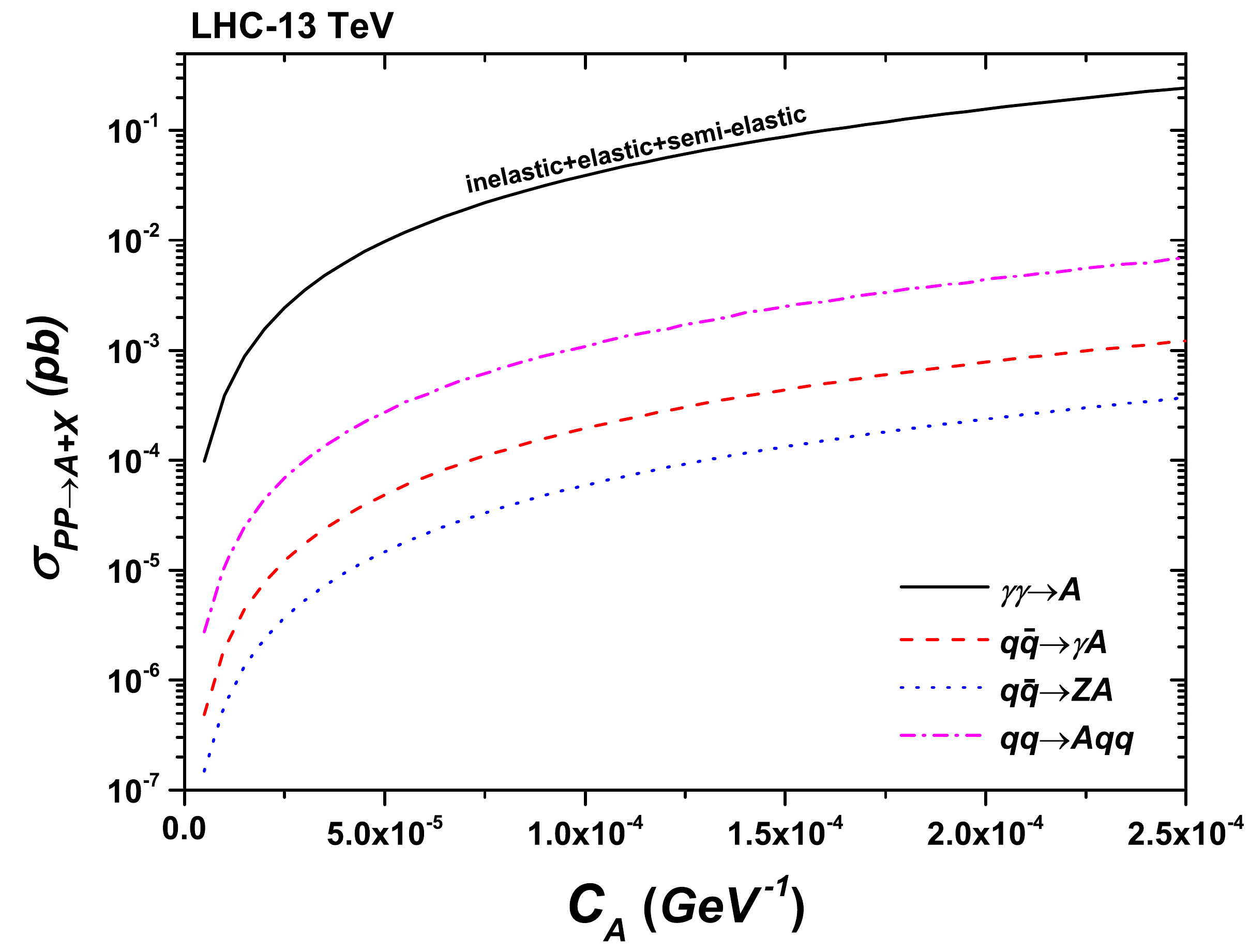}
	\caption{Dependence of the leading order production cross sections of the heavy axion on the effective coupling $C_{{\cal A}}$ at 13 TeV LHC.}
	\label{crossplot}
\end{figure}
In Fig. \ref{crossplot}, we present the dependence of the leading order production cross sections of the heavy axion on the effective coupling $C_{{\cal A}}$ at 13 TeV LHC. It can be seen that the production rate of the photon fusion is about ${\cal O}(10)$ times larger than that of the $ Z $ vector boson fusion. For example, when $C_{\cal A}=2.5\times 10^{-4}$, the cross sections of the photon fusion and the $ Z $ vector boson fusion can reach about 0.24 pb and 0.0046 pb, respectively. Also, due to the hierarchy between the couplings $C_{\gamma\gamma}$, $C_{\gamma Z}$ and $C_{ZZ}$ (c.f. Eq.(\ref{2.10})), the associated production of ${\cal A}\gamma$ has a much larger cross section than that of ${\cal A}Z$, which can reach about 1.3 fb for $C_{\cal A}=2.5\times 10^{-4}$ at 13 TeV LHC. So if the heavy axion dominantly decays to the diphoton, there will be a sizable number of three photon events at the future LHC searches. The rare decay $Z \to \gamma\gamma\gamma$ can also be induced, but its branching ratio is found to be less than $10^{-12}$, which is hardly observable at future colliders such as FCC-ee ($\sim 10^{12}$ events of $Z$) or CEPC ($\sim 10^{10}$ events of $Z$).

Note that if the heavy axion only has the couplings with $\gamma\gamma$, $Z\gamma$ and $ZZ$, the production cross sections and the decay widths of the heavy axion will be linearly correlated, which will lead to a tension between the predicted total width and the value $\Gamma \sim 45$ GeV, favoured by ATLAS. However, in our model, the heavy axion usually can have other decay modes. The corresponding decay partial width $\Delta\Gamma$ depends on the specific representation of the extra quarks, their masses and couplings. So we treat $\Delta\Gamma$ as a free parameter in our model. Then, the branching ratio of ${\cal A} \to \gamma\gamma$ can be computed by,
\begin{equation}
Br({\cal A} \to \gamma \gamma)= \frac{\Gamma_{\gamma\gamma}}{\Gamma_{\gamma\gamma}+\Gamma_{Z\gamma}+\Gamma_{ZZ}+\Delta\Gamma}
\end{equation}
where in the limit $m_{{\cal A}} \gg m_Z$,
\begin{equation}
\Gamma_{\gamma\gamma}=\frac{C_{\gamma \gamma}^2 m^3_\mathcal{A}}{4 \pi}, \quad \Gamma_{Z\gamma}=\frac{C_{Z \gamma}^2 m^3_\mathcal{A}}{8 \pi}, \quad \Gamma_{ZZ}=\frac{C_{ZZ}^2 m^3_\mathcal{A}}{4 \pi}.
\end{equation}
Due to $\Delta\Gamma\geq 0$, we can have the upper limit on the $Br({\cal A} \to \gamma\gamma)\simeq 61\%$, which in turn provides a lower limit on the effective coupling $ C_\mathcal{A} $. On the other hand, a very wide width of $\cal A$ is not favored by the LHC observation. We require the total width $\Gamma_{tot} \leqslant 45$ GeV in our following calculations.

\begin{figure}
	\centering
	\includegraphics[width=0.7\textwidth]{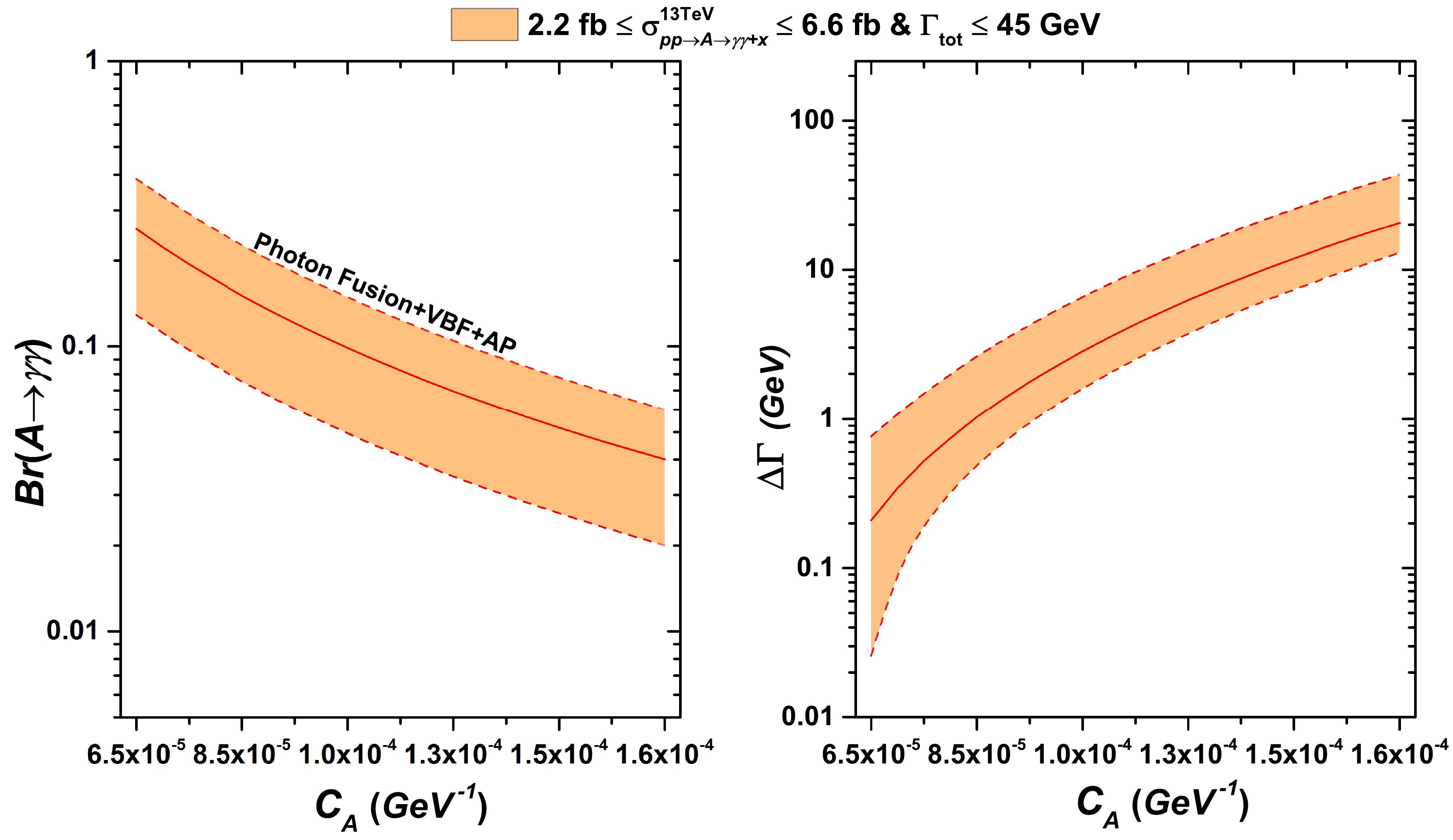}
	\caption{Contour plots of the branching ratio of ${\cal A} \to \gamma\gamma$ and the exotic decay width $\Delta\Gamma$ versus the effective coupling $C_{\cal A}$ within $2\sigma$ range of diphoton excess in Eq.(\ref{2.10}).}
	\label{aaA}
\end{figure}
In Fig. \ref{aaA}, we perform a fit of the branching ratio of ${\cal A} \to \gamma\gamma$, the exotic decay width $\Delta\Gamma$ and the effective coupling $C_{\cal A}$ to the observed number of excess events within $2\sigma$ range (cf. Eq. (\ref{excess})) under the total width bound $\Gamma_{tot}\leqslant 45$ GeV. From Fig. \ref{aaA}, we can see that when the coupling $C_{\cal A}$ becomes large, the branching ratio $Br({\cal A} \to \gamma\gamma)$ should be smaller to suppress the cross section of $pp \to {\cal A}+X$. In addition, the total width bound produces a upper limit on the coupling $C_{\cal A}<1.65 \times 10^{-4}$ GeV$ ^{-1} $ and a lower limit is provided by the upper limit of the branching ratio, $C_{\cal A}>6.5 \times 10^{-5}$ GeV$ ^{-1} $. We can also see that a small value of $C_{\cal A}$ is easier to satisfy the current experemental constraints and achieve the diphoton enhancement by a large branching ratio of ${\cal A} \to \gamma\gamma$.

The range of the effective coupling $ C_{\mathcal{A}} $ consistent with the required cross section is found to be $1.6\times 10^{-4}$ GeV$^{-1}\gtrsim C_{{\cal A}} \gtrsim 6.5\times 10^{-5}$ GeV$^{-1}$. Using Eq. (\ref{2.7},\ref{2.9}) and the values derived in Table 1, we can determine which high colour representation $ \mathcal{R} $ of $ \mathcal{Q} $ is most compatible with the allowed coupling range. We find that $ d(\mathcal{R})=6$ and $ 8 $ are ruled out by the predicted mass of the quark which violates $ m_{Q}<\mathcal{F}_{\mathcal{A}} $. For the $ d(\mathcal{R})=10  $ case we find $1.1\times 10^{-5} y_{Q}^2 $ GeV$^{-1} \gtrsim C_{\mathcal{A}}\gtrsim 1.6\times 10^{-6} y_{Q}^2$ GeV$^{-1}$, where the upper limit is due to the requirement of $ m_{Q}<\mathcal{F}_{\mathcal{A}} $. The $ d(\mathcal{R})=15  $ scenario can accommodate the coupling range $1.7\times 10^{-6} y_{Q}^2$ GeV$^{-1}\gtrsim C_{\mathcal{A}}\gtrsim 1.5\times 10^{-10} y_{Q}^2 $ GeV$^{-1}$. Each of these scenarios has a maximum effective coupling of approximately $\sim 10^{-6} y_{Q}^2 $, comparing this to Fig. 2, we see that consistency with observation tends to favour narrow width resonances for hypercharges of $ y_{Q}^2 \sim \mathcal{O}(10)$, within the accuracy of our approximations.

\section{Conclusion}
We have proposed a new model to account for the 750 GeV resonance that has been observed in the early LHC Run 2 data. This resonance can be accounted for by a 750 GeV heavy composite axion which results from the condensation of hypothetical quarks in a high-colour representation of conventional QCD. The axion mass and its coupling to two photons can be computed using the colour charge, hypercharge and mass of the new quarks, which can then be matched to the cross-section required by the LHC Run 2 analysis. The axion is predominantly produced via photon fusion which is followed by $ Z $ vector boson fusion and associated production at the LHC. We find that the total diphoton cross section of the axion can be fitted to the observed excess. Combining the requirements of the diphoton excess events and total width ($\Gamma_{tot} \leqslant 45$ GeV), we can obtain the allowed values of the effective coupling which are in the range $1.6\times 10^{-4}$ GeV$^{-1}\gtrsim C_{{\cal A}} \gtrsim 6.5\times 10^{-5}$ GeV$^{-1}$. This range can be satisfied in the $ d(\mathcal{R})= 10 $ and $ 15 $ cases within this model. Although, to obtain consistency between the model and allowed range we tend to require a narrow width resonance and $ y_{Q}^2\sim \mathcal{O}(10) $, within the accuracy of our approximations. 

We also notice that the associated production of $q\bar{q} \to {\cal A}\gamma$ can potentially produce a sizeable number of three photon events at future LHC. The rare decay $Z \to \gamma\gamma\gamma$ is also studied but is found too small to be probed at the LHCand $ e^{+}  e^{-} $ colliders.

\paragraph{Acknowledgement.} This work was partially supported by the Australian Research Council. AK was also supported in part by the Shota Rustaveli National Science Foundation (DI/12/6-200/13) and Lei Wu was also supported in part by the National Natural Science Foundation of China (NNSFC) under grants No. 11275057, 11305049, 11375001, 11405047, 11135003, 11275245, by Specialised Research Fund for the Doctoral Program of Higher Education under Grant No. 20134104120002; by the Open Project Program of State Key Laboratory of Theoretical Physics, Institute of Theoretical Physics, Chinese Academy of Sciences, P.R. China (No.Y5KF121CJ1).

\end{document}